\documentclass[12pt]{article}

\usepackage{amsmath}
\usepackage{graphicx}
\usepackage{enumerate}
\usepackage{natbib}
\usepackage{url} 

\usepackage{amssymb}
\usepackage{amsthm}
\usepackage{authblk}
\usepackage{bm}
\usepackage{epstopdf}
\usepackage{epsfig}
\usepackage{rotating}
\usepackage[margin=1in]{geometry}      
\usepackage{caption}
\usepackage{pgfplots}
\usepackage{tikz}
\usepackage{marginnote} 
\usepackage{algorithm}
\usepackage{setspace}
\usepackage{algpseudocode}
\usepackage{subfigure}
\usepackage{makecell}
\usepackage{lscape}
\usetikzlibrary{quotes,angles}
\usepackage{nameref}
\usepackage[hyphenbreaks]{breakurl}
\usepackage{soul}
\usepackage{xcolor}

\definecolor{lightgreen}{rgb}{0.56, 0.93, 0.56}
\definecolor{lightfuchsiapink}{rgb}{0.98, 0.52, 0.9}

\newcommand{\revise}[1]{{\color{black} #1}}

\newcommand{\bc}{\mathbf{c}}
\newcommand{\bu}{\mathbf{u}}
\newcommand{\bv}{\mathbf{v}}
\newcommand{\bw}{\mathbf{w}}
\newcommand{\bx}{\mathbf{x}}
\newcommand{\by}{\mathbf{y}}
\newcommand{\bz}{\mathbf{z}}
\newcommand{\bK}{\mathbf{K}}
\newcommand{\bV}{\mathbf{V}}
\newcommand{\bW}{\mathbf{W}}
\newcommand{\bX}{\mathbf{X}}
\newcommand{\bZ}{\mathbf{Z}}

\newcommand{\bzero}{\mathbf{0}}
\newcommand{\bone}{\mathbf{1}}
\newcommand{\beps}{\bm{\epsilon}}
\newcommand{\bgam}{\bm{\gamma}}

\pgfplotsset{compat=1.12}

\title{Mixed Effect Modeling and Variable Selection for Quantile Regression}
\author{Haim Y. Bar\thanks{
    haim.bar@uconn.edu. The authors gratefully acknowledge the following funding support: Prof. Bar's research was supported by NSF-DMS
1612625. Professor Booth's research was partially supported by NSF-DMS 1611893. Professor Wells' research was partially supported by NSF-DMS 1611893, and NIH grant U19 AI111143.}\\
    315 Philip E. Austin Building
Department of Statistics,
University of Connecticut
Storrs, CT, 06269-4120, USA.\\
    and \\
    James G. Booth and Martin T. Wells\\
    Department of Statistics and Data Science, Cornell University, \\ Ithaca NY, 14853, USA.}
\pgfplotsset{compat=1.12}

\begin{document}

\maketitle
\abstract{
It is known that the estimating equations for quantile regression (QR) can be 
solved using an EM algorithm in which the M-step is computed via
weighted least squares, with weights computed at the E-step as the
expectation of independent generalized inverse-Gaussian variables.
This fact is exploited here to extend QR to allow for random
effects in the linear predictor. Convergence of the
  algorithm in this setting is established by showing that it is a generalized alternating
minimization (GAM) procedure. Another modification of the EM algorithm
also allows us to adapt a recently proposed method for variable
selection in mean regression models to the QR setting. 
Simulations show the resulting method significantly outperforms
variable selection in QR models using the lasso penalty. Applications to real data include
a frailty QR analysis of hospital stays, and variable selection for age at
onset of lung cancer and for riboflavin production rate using
high-dimensional gene expression arrays for prediction. 
}


\section{Introduction}\label{sec.introduction}
Quantile regression (QR) is used to predict how percentiles of a quantitative response variable, $y$,
change with some predictors, $\mathbf{x}=(x_1,\ldots,x_P)$ and, as a
consequence, offers an approach for examining how covariates influence
the location, scale, and shape of the response distribution. For
example, it is reasonable to investigate whether covariates
will have different effects in the tails of the distribution than in
the center. A nice illustration of this point is provided by
\cite{koen:hall:2001} who fit a QR model to study the determinants
of infant birth weight and show, for example, that although boys are
heavier than girls on average ...\textit{the disparity is much smaller in the
lower quantiles of the distribution and considerably larger ... in the
upper tail of the distribution.}

In what follows it is assumed that the $q$-th quantile of $y$ is
related to $\mathbf{x}$ via a linear model. Specifically,
$y_1,\ldots,y_n$ are independently distributed as
$P(y_i<y)=F(y-\mathbf{x}_i \bm{\beta}_q)$, where $\bm{\beta}_q$ is a $P \times 1$ 
vector of unknown coefficients which depends on $q\in(0,1)$. 
(Henceforth $\mathbf{x}_i \bm\beta_q$ will include an intercept, $\beta_{0,q}$.)

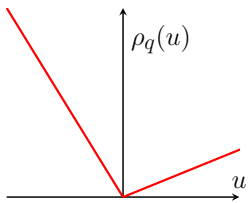
\begin{figure}[b!]
    \begin{center}
    \begin{tikzpicture}[scale=0.7 ]
    \pgfmathsetmacro{\a}{0.1}
    \pgfmathsetmacro{\b}{1} 
    \pgfplotsset{ticks=none}
      \begin{axis}[axis lines=middle,xmin=-0.75,xmax=0.75,ymin=0,ymax=0.75*0.8, thick, width=6cm,
        xlabel=$\scriptstyle u$,
        ylabel=$\scriptstyle \rho_q(u)$,
        x label style={at={(1,0.15)},anchor=north,font=\LARGE},
        y label style={at={(0.5,0.95)},font=\LARGE},
       no marks,
       samples=100
            ]
        \coordinate (O) at (0,0);
    \addplot+[red,domain=-0.75:0,very thick] {-0.8*x};
    \addplot+[red,domain=0:3,very thick] {0.2*x};
          \end{axis}
    \end{tikzpicture}
    \end{center}      
    \caption{The `check' loss function ($q=0.2$)\label{checkloss}.}
    \end{figure}

In the QR framework, estimation of the regression parameters (for
a specific quantile of interest, $q$) is done by solving an estimating equation involving the `check' loss function (see Figure \ref{checkloss}):
\begin{equation}\label{objectiveFunction}
 \hat{\bm{\beta}}_q=\underset{\bm{\beta}}{\arg\min}\sum_{i=1}^n\rho_q\left(y_i-\mathbf{x}_i \bm{\beta}\right)
\end{equation}
where
\begin{equation}
 \rho_q(u)=u\cdot(q-1_{[u<0]})\,,
\end{equation}
and $1_{[u<0]}$ is the indicator function that equals 1 when the argument, $u$, is negative.
Equation (\ref{objectiveFunction}) does not lead to a closed-form formula for $\bm\beta_q$,
but an efficient numerical solution using linear programming methods is available. 
A comprehensive review of QR is given in \cite{koen:2005}.

The optimization problem (\ref{objectiveFunction}) is equivalent to maximum likelihood estimation
under the assumption that the errors in the linear model,
$y_i = \mathbf{x}_i \bm\beta_q + u_i$,
are independent draws from an asymmetric Laplace distribution (ALD) with
density given by
\begin{eqnarray}
\label{ALDensity}
  h_q(u|\alpha)=q(1-q)\exp\left\{ -\rho_q( u/\alpha ) \right\}/\alpha\,.
\end{eqnarray}
\cite{yu:moye:2001}, for example, exploit this fact to develop a fully
Bayesian approach to QR.
In a Bayesian approach to quantile regression, \cite{kozumi2011gibbs} posited the ALD as the working likelihood to perform the inference.
 It is also well known that the asymmetric Laplace
distribution can be derived as a scale-mixture of distributions; see, for example, \cite{gera:bott:2014}, \cite{Galarza:2017}, and \cite{gala:band:lach:2017}. 

\cite{zhou:ni:li:2014} show that the optimization problem in
(\ref{objectiveFunction}) can be solved using an EM
algorithm that involves fitting a normal theory linear
model via weighted least squares at the M-step, and updating the
weights at the E-step. Furthermore, the EM algorithm is
  operationally equivalent to
the majorize-minimization (MM) algorithm described in
\cite{HunterLange:2000}.
The key contributions in this paper are two modifications of the EM
algorithm that allow for (i) the extension of QR to
the mixed effects setting; and (ii) to implement a 
variable selection algorithm described in \cite{SEMMS} in the QR context.

Other methods for fitting quantile regression mixed models
using the asymmetric Laplace distribution
have been proposed; see, for example, \cite{gera:bott:2007},
\cite{gera:bott:2014}, and \cite{gala:band:lach:2017} who use
an EM-type algorithm combined with numerical quadrature for fitting
the QR model. Similarly, \cite{Taddy:2010} proposed a Bayesian
approach based on a scale-mixture of normal distributions. However, these methods lack the
computational efficiency required to address the high-dimensional
variable selection problems we consider.

Several authors proposed a regularization approach to variable
selection in the quantile regression setting, including \cite{belloni2011} using $\ell_1$ penalty, and \cite{Yi:2017} using an elastic-net penalty. An implementation of penalized quantile regression is also available in recent \cite{rqPen}. Bayesian regularization approaches relying on the scale-mixture model include \cite{ParkCasella:2008} and \cite{LiXiLin:2004}.

The paper proceeds as follows. In Section \ref{sec.model} we describe
the EM algorithm for solving the QR optimization problem in
(\ref{objectiveFunction}). In Section \ref{sec.mixedmodels} we propose
an extension to incorporate mixed predictors in QR models 
and  in Section \ref{sec.VariableSelection} we develop a variable selection algorithm for QR.
Section \ref{sec.simulation} summarizes a simulation study, and in Section
\ref{sec.casestudy} we apply our method to three data sets, including
a frailty QR analysis of hospital stays, and variable selection for
age of onset of lung cancer and for riboflavin production rate using
high dimensional gene expression arrays for prediction.
Section \ref{sec.discussion} contains a brief conclusion and discussion of future work.

\section{An EM Algorithm for Quantile Regression}\label{sec.model}
Following \cite{zhou:ni:li:2014}, suppose that mutually independent
paired random variables, $(u_i,w_i)$, where
$u_i=y_i-\mathbf{x}_i \bm\beta_q$, $i=1,\ldots,n$, have joint density
$$p(u_i,w_i | \bm\beta_q)= \frac{2q(1-q)}{\sqrt{2\pi w_i}}\exp\left\{ -\frac{[u_i-(1-2q)w_i]^2 }{ 2w_i } \right\}\exp\{ -2q(1-q)w_i \}\,.$$
Thus, the $w_i$ are iid latent exponential variables with rate
$2q(1-q)$, and conditional on $w_i$, $u_i$ is normally distributed with mean
$(1-2q)w_i$ and variance $w_i$. The latter implies that, conditionally
on $\mathbf{w}=(w_1,\ldots,w_n)^T$, the response
vector $\mathbf{y}=(y_1,\ldots,y_n)^T$ follows a linear model with
mean $\mathbf{X}\bm\beta_q+(1-2q)\mathbf{w}$ and variance-covariance
matrix, $\mathbf{W}=\mbox{diag}(\mathbf{w})$. It is straightforward to verify that
the marginal density of $u_i$ is of the form (\ref{ALDensity}) with
$\alpha=1/2$, and the conditional density of $w_i^{-1}$ given $u_i$ is inverse
Gaussian, $\mathcal{IG}(|u_i|^{-1},1)$.

These results imply that the optimization problem (\ref{objectiveFunction}) can be
solved using the EM algorithm \citep{EM}, in which the latent
$w_i$'s are the missing data. 
The complete data log-likelihood, ignoring terms not involving
$u_i$ and hence also $\bm\beta_q$, is given by
$$ \sum_{i=1}^n\left( -\frac{u_i^2}{2w_i} + (1-2q)u_i \right) \,. $$
Let $\hat{\bm\beta}_q'$ denote the value of $\bm\beta_q$ at the current
iteration of EM. Then it follows that the E-step of the EM algorithm involves replacing
$w_i^{-1}$ by its conditional mean $|u_i|^{-1}$ evaluated at
$\hat{\bm\beta}'_q$, and the M-step update via
weighted least squares is
\begin{eqnarray*}
  \hat{\bm\beta}_q=(\mathbf{X}^T\hat{\mathbf{W}}^{-1}\mathbf{X})^{-1}\mathbf{X}^T\hat{\mathbf{W}}^{-1}\left[\mathbf{y}-(1-2q)\hat{\mathbf{w}}\right]\,.
\end{eqnarray*}
Implementation details are summarized in Algorithm \ref{algorithm}, where $\ell$ is the conditional
log-likelihood of $\by|\bw$, and $\bm\beta'_q$ is initialized using the ordinary least square
estimator.

\begin{algorithm}
\caption{The Quantile Regression EM (QREM) algorithm}\label{algorithm}
\begin{algorithmic}[1]
\State Initialize $\epsilon>0$, $\delta \leftarrow 2\epsilon$
\State Initialize $\hat{\bm\beta}^{'}_q \leftarrow (\bX^T\bX)^{-1}\bX^T\by$
\While{$\delta > \epsilon$}
\State E-Step: $\hat{w}_i^{-1} \leftarrow |y_i-\mathbf{x}_i \hat{\bm\beta}_q^{'}|^{-1}$
\State M-Step: $\hat{\bm\beta}_q \leftarrow (\bX^T\hat\bW^{-1}\bX)^{-1}\bX^T\hat\bW^{-1}\left[\by-(1-2q)\hat{\bw}\right]$
\State $\delta \leftarrow |\ell(\by|\bX, \bw,\hat{\bm\beta}^{'}_q)-\ell(\by|\bX, \bw,\hat{\bm\beta}_q)|$
\State $\hat{\bm\beta}^{'}_q\leftarrow\hat{\bm\beta}_q$
\EndWhile
\end{algorithmic}
\end{algorithm}

We use the same approach as in \cite{zhou:ni:li:2014} to deal with the case of  zero residuals in the E-step.
Since Algorithm \ref{algorithm} solves the QR optimization
problem (\ref{objectiveFunction}), we can invoke known results concerning the properties
of the solution. In particular, under suitable regularity conditions (see Appendix \ref{consistencyConditions}) the estimator is consistent and
asymptotically normal with asymptotic variance-covariance matrix given
by
\begin{eqnarray*}
  \mbox{a.var}(\hat{\bm\beta}_q)=\frac{q(1-q)}{f(0)^2}(\bX^T\bX)^{-1}\,,
\end{eqnarray*}
where $f(0)$ is the density of $u_i$ at zero \citep{rupp:carr:1980,koen:2005}. 

If the model is correct, $E[\mbox{sign}(y_i-\bx_i \bm\beta_q)]=1-2q$,
for $i=1,\ldots,n$. Hence the binary variables
$c_i=\mbox{sign}(y_i-\bx_i \bm\beta_q)-(1-2q)$ have mean zero and
$E(\bX^T\bc)=0$, where $\bc=(c_1,\ldots,c_n)^T$. Thus, $\bc$ is
uncorrelated with every predictor. This suggests that the pattern of
predictors associated with $c_i$ values equal to $2q$ should be similar to the pattern of
predictors associated with $c_i$ values equal to $2q-2$. In fact if
the vector $\bc$ is replaced by its estimate,
$\hat\bc=\mbox{sign}(\by-\bX \hat{\bm\beta}_q)-(1-2q)\bone\,,$ 
then $\bX^T\hat\bc\equiv 0$. 
(A proof of this result is given in Appendix \ref{propUncorr}.)
These facts can be used to produce diagnostic plots as follows. 
Let $A=\left\{i : \hat{c}_i = 2q \right\}$ and $B=\left\{i : \hat{c}_i = 2q-2\right\}$ 
be the sets of observations whose responses are above and below the quantile regression
fitted values respectively. Suppose that the $j$th column of $\bX$ is a continuous predictor.
If the regression model is correct
we expect the Q-Q plot of $x_{ij}$ for $i \in A$ versus $x_{ij}$ for $i \in B$ 
to lie close to a $45^\circ$ line. Deviation from this pattern
suggests a deficiency in the model. 
Use of these diagnostic Q-Q plots is demonstrated
in Sections \ref{sec.simulation} and \ref{sec.casestudy}. 
For categorical predictors, if the model is correct, the expected
proportion of responses with $c_i=2q$ is $1-q$ for each category level.

So far we have not assumed a fully specified parametric model for the
data generating mechanism. The distributional assumptions for
$(\bu,\bw)$ were merely a device for creating an algorithm for solving
the QR optimization problem (\ref{objectiveFunction}).
However, by analogy with the classical linear model framework, a natural measure of
goodness of fit is $G=2n\bar{G}_{\rho_q}$, where
$\bar{G}_{\rho_q}=n^{-1}\sum_{i=1}^n\rho_q(u_i)$,  is the mean check
loss error, so that
\begin{equation}\label{GOF}
G=-\log(\mbox{maximum marginal likelihood of $u$})+\mbox{constant}(q)\,, 
\end{equation}
under an ALD assumption for the QR model errors. An obvious
modification is to use AIC$=2G+2P$ for model comparisons.

\section{Extension to Mixed Models}\label{sec.mixedmodels}
The classical linear mixed model is a widely used approach for
modeling the mean as a function of predictors 
in applications involving dependent responses. The generic form of
the linear mixed model is 
\begin{equation}
\by = \bX \bm\beta + \bZ \bv + \bm\epsilon\,,
\end{equation}
where the dependence is captured by random effect parameters,
$\bv$, having a zero-mean multivariate normal distribution with 
(often structured) variance-covariance matrix $\bK$ 
\citep{pinh:bate:2000,sear:case:mccu:1992}. 
The random error vector, $\bm\epsilon$, is assumed to be normally
distributed and independent of the random effects. 

\cite{koen:2005} refers to longitudinal studies and says that
`...\textit{many of the tricks of the trade developed for Gaussian random effects models are
no longer directly applicable in the context of quantile regression}.'
This is due to the fact that the regression estimates are not obtained via the
convenient and tractable least squares estimation method.  In particular, for longitudinal regression the non-smoothness of the objective function complicates the analysis of the QR estimator. 
However, with the mixture representation approach, by modifying the M-step of the EM algorithm described in the previous
section, it is possible to extend the classical mixed model to the QR
setting. Specifically, let 
$\mathbf{u}=\by-\bX \bm\beta_q-\bZ \bv$ and 
suppose that, conditionally on $\bv$, the paired random variables,
$(u_i,w_i)$, $i=1,\ldots,n$, are iid with joint distribution 
\begin{eqnarray*}
 w_i|\bv &\sim& Exp(2q(1-q)) \\
 u_i|w_i, \bv &\sim& N((1-2q)w_i,w_i).
\end{eqnarray*}
As a result, $w_i^{-1}|u_i, \bv \sim IG(|u_i|^{-1}, 1)$,
$i=1,\ldots,n$, are conditionally independent given $\bv$. 
As in the classical case the error vector $\bu$ is assumed to be
independent of the random effects vector $\bv$ but, unlike in the
classical case, the error vector is not assumed to be normally
distributed. Implementation details are summarized in
Algorithm \ref{algorithm2}. 

\begin{algorithm}
\setstretch{1}
\caption{The Extended Quantile Regression EM (EQREM) algorithm}\label{algorithm2}
\begin{algorithmic}[1]
\State Initialize $\epsilon>0$, $\delta = 2\epsilon$
\State Initialize $\hat{\bm\beta}^{'}_q\leftarrow(\bX^T\bX)^{-1}\bX^T\by$, $\hat\bv^{'}\leftarrow\bzero$
\While{$\delta > \epsilon$}
\State F-Step: $\hat{w}_i^{-1}\leftarrow |y_i-\bx_i \hat{\bm\beta}_q^{'}-\bz_i \hat{\bv}^{'}|^{-1}$
\State B-Step (REML): $\hat{\bK}\leftarrow$ REML for $\bK$ given $\hat\bw$
\State B-Step (BLUE): $\hat{\bm\beta}_q\leftarrow$ BLUE for $\bm\beta_q$ given $\hat\bw$
\State B-Step (BLUP): $\hat\bv\leftarrow$ BLUP for $\bv$ given $\hat\bw$
\State $\delta=|\ell(\by|\bv,\bX, \bw,\hat{\bm\beta}^{'}_q)-\ell(\by|\bv,\bX, \bw,\hat{\bm\beta}_q)|$
\State $\hat{\bm\beta}^{'}_q\leftarrow\hat{\bm\beta}_q$
\EndWhile
\end{algorithmic}
\end{algorithm}

While Algorithm 1 can be implemented using existing tools such as the
\texttt{lm()} function \citep{R:2018} for fixed effect model which were built for
the mean-model setting with normal errors,
Algorithm \ref{algorithm2} requires methods for mixed models such as
\texttt{lmer()} \citep{lme4:2015}, or equivalent
procedures in other languages such as SAS and Stata. This also
facilitates the estimation of the matrix $\bK$ when fitting mixed
models. 

The key difference relative to the fixed effect model is that in order
for the components of $\bw^{-1}$ to be independent inverse Gaussian
variates, we have to condition not only on $\bu$, but also on $\bv$.
To justify plugging in the best linear unbiased predictor (BLUP) for
$\bv$, we rely on results in \cite{Gunawardana:2005} and their
generalization of the EM algorithm, called GAM (Generalized
Alternating Minimization).  Like the EM algorithm, GAM consists of two
steps. The `backward' step generalizes the M-step, and the `Forward'
step generalizes the E-step.  In this case $\bm\beta$ and $\bK$ are the parameters and
$\bw$ and $\bv$ are the missing data.  Because
the complete data likelihood can be written in closed-form, the
backward step in our case is identical to the M-step, and
$\hat{\bm\beta}_q$ and $\hat{\bK}$ are the maximum likelihood estimators
given the current imputed value of $\bw$.

Regarding the Forward step, \cite{Gunawardana:2005} refer to a
probability distribution function, $Q_C$, as `desired' if it has the
properties that (i) the maximum likelihood is obtained with the
observed data, and (ii) it reduces the Kullback-Leibler divergence
relative to the previous iteration.  They denote the set of desired
distribution functions by $\mathcal{D}$.  The objective in the forward
step is to find $Q_C\in\mathcal{D}$ such that
\begin{equation}\label{KL}
 D_{KL}(Q_C^{(t+1)}||P_{C}(\hat{\bm\beta}^{(t)},\hat{\bK}^{(t)})) \le D_{KL}(Q_C^{(t)}||P_{C}(\hat{\bm\beta}^{(t)},\hat{\bK}^{(t)}))\,,
\end{equation}
where $P_{C}(\hat{\bm\beta}^{(t)},\hat{\bK}^{(t)})$ is the member of
the parametric family of the complete data likelihood, evaluated at
the MLE's after iteration $t$.  Note that the objective in the EM
algorithm is to find a $Q_C$ which minimizes the KL divergence on the
left-hand side of (\ref{KL}), while in order to guarantee the
convergence of a GAM procedure it is sufficient to find any desired
distribution which reduces it.

While finding a simultaneous update for $\bv$ and $\bw$ may be
intractable, the Forward step can be performed in two steps, and the
convergence of GAM will still hold.  Let
$\mathcal{D}_{\bv|\bw}\subset\mathcal{D}$ be the set of desired
distributions which satisfy (\ref{KL}) while holding $\bw$ fixed at
their current value. Clearly, the BLUP not only reduces the KL
divergence, but it actually minimizes it, making the BLUP the optimal
next estimate for $\bv$, given $\bw$.  Now, let
$\mathcal{D}_{\bw|\bv}\subset\mathcal{D}$ be the set of desired
distributions which satisfy (\ref{KL}) while holding $\bv$
fixed at their current value (the BLUP). Then, given $\bv$ and
$u_i$, the best update for $w_i^{-1}$ is $|u_i|^{-1}$, as we
have shown previously in the fixed effect model.
This two-step approach is valid because (i) any distribution function
obtained from a projection of $\mathcal{D}$ to a subspace is also a
valid candidate for $Q_C^{(t+1)}$ when using $\mathcal{D}$, because
GAM does not require finding the minimizer -- just an improvement with
respect to the previous iteration; and (ii) with each of the two
projections into $\mathcal{D}_{\bv|\bw}$ and
$\mathcal{D}_{\bw|\bv}$ the conditions of the GAM convergence
theorem in \cite{Gunawardana:2005} hold.

\section{Variable Selection in Quantile Regression}\label{sec.VariableSelection}
In this section we consider the setting where the number of
predictors, $P$, is large, possibly
much larger than $n$, but it is assumed that only a small number of
columns of $\bX$ are actually related to the $q$th quantile of the response. 

\revise{In the frequentist high-dimensional literature there have been a numerous penalized likelihood approaches used for variable selection.  
The most popular method is the LASSO \citep{tibshirani1996regression}, which uses the  penalty function $p_\lambda (\beta_j) = \lambda|\beta_j |$. Besides the LASSO and its many variants, other popular choices for $p_\lambda (\beta_j) $ include non-concave penalty functions, such as the smoothly clipped absolute deviation (SCAD) penalty \cite{fan2001variable} and the minimax concave penalty (MCP) \cite{zhang2010nearly}. All of these penalties force some coefficients to zero, thus enabling to perform variable selection.  SCAD and MCP type penalties also mitigate the estimation bias of the LASSO.  \cite{rqPen} and \cite{sottile2020penalized} give suites of penalization methods for variable selection in quantile regression.}

\revise{In the Bayesian framework, variable selection for linear models arise directly from probabilistic modelling of the underlying parameter sparsity and is frequently carried out through assigning a spike-and-slab prior on the coefficients of interest. The spike-and-slab prior was first introduced by \cite{mitchell1988bayesian}.  The point-mass spike-and-slab type prior is often considered theoretically optimal for sparse Bayesian problems  \citep{castillo2012needles, johnstone2004needles, ishwaran2005spike, zhang2010generalized}.  However, typically in high dimensional estimation settings exploring the full posterior using point-mass spike-and-slab priors can be computationally onerous because of the combinatorial complexity of updating the discrete components. }

\revise{Our hierarchical mixture prior is similar to a spike-and-slab model. The main difference is our choice of a three-way mixture model, in which there are two non-null components, rather than one. Compared with the spike-and-slab approach, our model offers advantages.  The two-component mixture assumes that the non-null distribution is symmetric, which implies a prior belief that the proportion of variables which are positively correlated with the response is the same as the proportion of predictors that are negatively correlated with the response, this may be an unreasonable assumption.  The non-null component in the two-component mixture has much of its mass around zero, which is counter-intuitive because it is assumed that variables in the non-null component have a non-zero effect. In contrast, the three-component model assigns a very small probability to non-null values near zero. Our mixture model also allows for the non-null components to be highly concentrated, which may be especially useful in situation where there is a single significant predictor. In addition, the three-component model allows us to borrow information across the two non-null components. See \cite{SEMMS} for more on the benefits of this class of  three-component mixture priors.}

We propose an algorithm which iterates between the following two steps 
until convergence is achieved: 
\begin{enumerate}
\item Select a set of candidate variables.\label{vsstep}
\item Fit a QR model using only the selected predictors.\label{qrstep}
\end{enumerate}

For the variable selection step (\ref{vsstep}) we use the empirical
Bayes approach to variable selection
from \cite{SEMMS}, which we describe here very briefly. The method is
applicable to any generalized linear model, and allows for two types
of predictors -- `locked in' variables that are always to be included
and a large number of `putative' predictors from which only a small 
subset are to be included.
For the purpose of performing variable selection in the quantile
regression setting, it is sufficient to focus on the special case in
which the responses are normal, although this does not require or imply an
assumption of normal errors in the QR model. To further simplify
the presentation we do not include the `locked in' variables in the following equations. 
Consider the model
\begin{eqnarray}\label{largePmodel}
  y_{i}= \beta_0+\sum_{k=1}^{P}x_{ik}\gamma _{k}v_{k}+\epsilon_i\,,
\end{eqnarray}
$i=1,\ldots,n$, where $\epsilon_i \stackrel{iid}{\sim} N(0,\sigma^2_\epsilon)$, $v_{k} \stackrel{iid}{\sim} N\left( \mu
  ,\sigma_v^{2}\right)$, and 
$\gamma _{k} \stackrel{iid}{\sim}\mbox{Mult}\left(-1,0,1; p_L,p_0,p_R\right)$,
and the random vectors, $\beps$, $\bv$ and $\bgam$, are
independent.  \revise{The three-way mixture model specification for $(v_{k}, \gamma _{k})$ for $k=1, \ldots, P$ is an extension of the two-way mixture spike-and-slab prior. } 

Variable selection with this mixture model is essentially a
classification problem, in which the goal is to identify for which
$k$, $\gamma_k\ne 0$. Estimation using the
Generalized Alternating Minimization (GAM) algorithmic framework of
\cite{Gunawardana:2005} involves maximizing a normal theory linear mixed model
likelihood and updating the latent $\left\{\gamma_k\right\}$ as their
posterior means after each iteration.

The method is implemented in an R package called SEMMS \citep{SEMMS} that
takes as input an initial set of variables, $\bV_0$ and outputs the
selected variables, $\bV_f$, where $f$ is the number of iterations
ultimately performed by the algorithm. At iteration $j$ the variable
set, $\bV_j$, is modified by adding or removing a single variable 
if doing so increases the log-likelihood 
by more than a predefined threshold $\delta>0$.
If no such variable exists, the algorithm terminates. 
The output from this variable selection algorithm is denoted by $SEMMS(\by,\bX, V_0)$.

The extension of SEMMS to the QR context is straightforward because
it can also be accomplished using a GAM algorithm that involves
fitting a normal theory linear 
 model at each iteration.
Details of the algorithm are given in Algorithm
\ref{algorithmLargeP}, in which $\bX_{[S]}$ denotes a
subset of the columns of the matrix $\bX$ indexed
by $S$. 

\begin{algorithm}
\setstretch{1}
\caption{Variable Selection for Quantile Regression}\label{algorithmLargeP}
\begin{algorithmic}[1]
\Require{$\by$ ($n$-vector, numeric), $\bX$ ($n\times P$ matrix, numeric), 
and $q\in(0,1)$}
\State Initialize $\epsilon>0$
\State $\ell\leftarrow 0$
\State $S\leftarrow$ an initial subset of  predictors for $Q_q(\by|\bX)$\label{init}
\Repeat
\State $\ell'  \leftarrow \ell$
\State $\hat{\bm\beta}_{q} \leftarrow EQREM(\by, \bX_{[S]},q)$\label{qrem}
\State $\mathbf{u}  \leftarrow \mathbf{y}-\mathbf{X}_{[S]} \hat{\bm\beta}_{q}$, for $i=1,\ldots,n$
\State $S'\leftarrow SEMMS(\by-(1-2q)|\bu|, \bX, S)$\label{ytilde}
\State $S\leftarrow S'$
\State $\ell \leftarrow -2\sum_{i=1}^n \rho_q(u_i)$
\Until $|\ell-\ell'| < \epsilon$
\Ensure $\hat{\bm\beta}_{q}$
\end{algorithmic}
\end{algorithm}

In the initialization step (\ref{init}) one can obtain $S(0)$ by
fitting $P$ one-at-a-time quantile regression models, and select the
$K$ most significant predictors for $Q_q(\by|\bX_{[i]})$, for some
small $K$.  Since this may be slow when $P$ is large, an alternative,
using an assumption that the effect of the predictors is additive, is
to divide the $P$ predictors into $\lceil P/m\rceil$, where $m<n$,
non-overlapping subsets, fit a quantile regression model with each
subset, and then pick the $K$ most significant predictors from all
$\lceil P/m\rceil$ fitted models. Another possibility is use the lasso
or any of its variants to do the initialization. In this case, as
noted by \cite{SEMMS}, Algorithm \ref{algorithmLargeP} is guaranteed to do at least as well as the
method used to initialize it because it yields a non-increasing Kullback-Leibler divergence.
 
Note that the convergence criterion is based on the logarithm of the
\textit{marginal} likelihood of $\bu$ under the assumption of ALD errors, defined as
$\ell = -2\sum_{i=1}^n \rho_q(u_i)$, which was used to define the
goodness-of-fit measure for the fitted quantile regression model in
(\ref{GOF}).  Thus, Algorithm \ref{algorithmLargeP} terminates when
the change in the log-likelihood (or equivalently, the improvement to
the goodness of fit) between consecutive iterations, is sufficiently
small.  

\medskip\noindent{\bf Proposition}:
Let $\by$ be a numeric vector, $\bX$ an $n\times P$ matrix where $P$ is large, and $q\in(0,1)$. 
Assume that only a small (but unknown) number of columns of $\bX$ are related to
the $q$th quantile of $\by$, so that
$Q_q(\by|\bX)=\bX_{[S]}\bm\beta_q$ for some subset of $L$ columns of
$\bX$ and an $L$-dimensional vector $\bm\beta_q$, where $L\ll P$.
Under these conditions Algorithm \ref{algorithmLargeP} is guaranteed
to converge.

\medskip
The proof of the convergence is similar to the one used in Section \ref{sec.mixedmodels} to show
 the convergence of the QREM algorithm
in the case of fitting a mixed effect quantile regression model, in the sense that 
estimating the parameters and latent variables in separate steps 
consisting of disjoint subsets of parameters and latent variables still falls
within the GAM framework of \cite{Gunawardana:2005} and thus convergence is assured, if, as is the case here, (i) in each step the Kullback-Leibler 
divergence relative to the previous iteration is reduced, and (ii) the parameter estimation is done via
maximum likelihood using the observed data.

Note that in general, when $P$ is large the posterior distribution obtained from (\ref{largePmodel})
can be multimodal, and hence, there can be more than one set of predictors which fit the data well.
\cite{SEMMS} recommend running the variable selection algorithm multiple times using the
so-called `randomized' version of  SEMMS, thus allowing users 
to obtain different, but similarly well-fitting models.

\clearpage

\section{Simulations}\label{sec.simulation}
Table \ref{sims} in Appendix \ref{sec.appendix} contains details
regarding 25 different simulation scenarios, performed in order to
assess the performance of Algorithms \ref{algorithm} and \ref{algorithm2} in terms of bias
and variance of regression parameter estimates.  In some simulations
the error variance does not satisfy the usual mean-model regression
assumptions, namely, being normally distributed and uncorrelated with
the predictors.  In scenarios 14-18 the error variances depend on a
predictor, and in 19-22 the error terms are sampled from a skewed
distribution (lognormal).  Table \ref{simsLargeP} in Appendix
\ref{sec.appendix} lists nine simulation scenarios in the large $P$
setting, designed to assess the performance of Algorithm \ref{algorithmLargeP}.

\begin{figure}[hb!]
  \begin{center}
 \includegraphics[ scale=0.8]{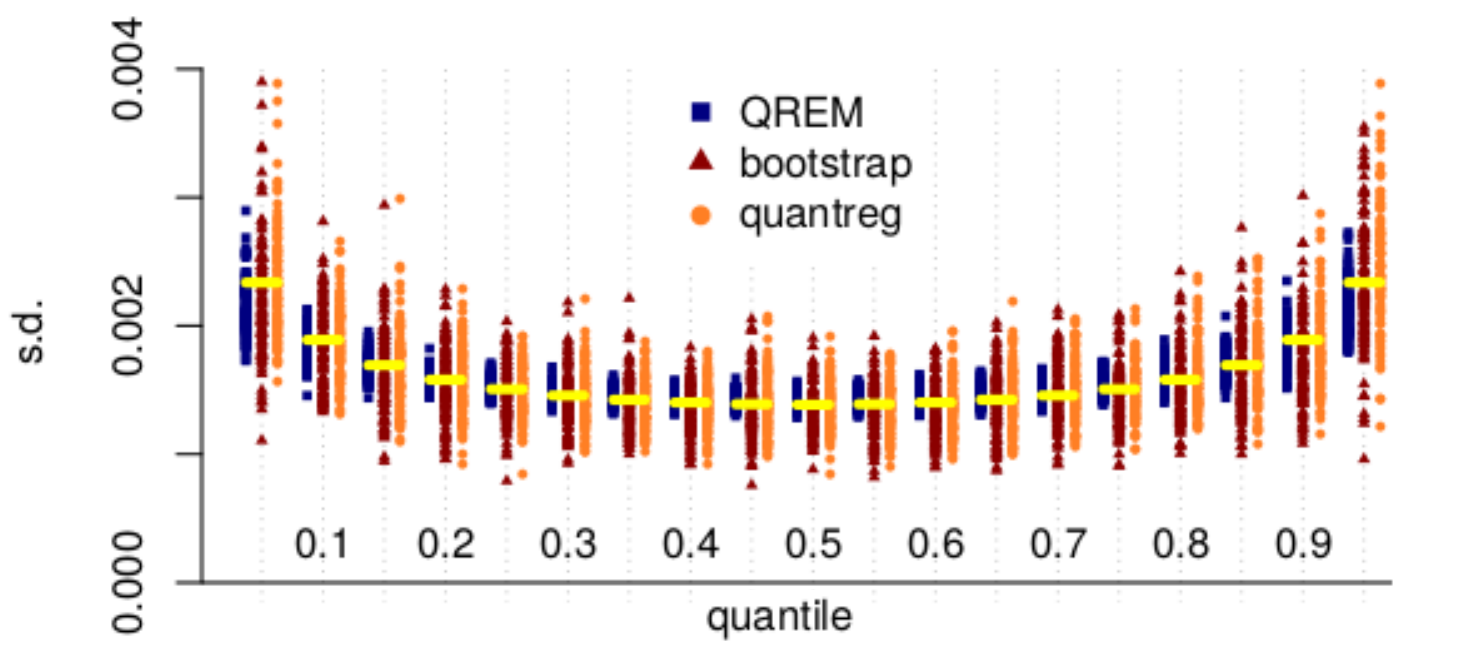}
\end{center}
\caption{QR simulation: $\hat\sigma_{\hat\beta_1}$ for $q\in\{0.05,0.1,\ldots,0.95\}$ for
$y\sim N(-3+x, 0.1^2)$.}\label{sdbeta1}
 \end{figure}
 
 For the scenarios listed in Table \ref{sec.appendix}\ref{sims}
 we compared the estimated regression coefficients with those obtained
 from the \texttt{quantreg} package
 \citep{quantreg:2018}, which uses a different estimation approach
 (namely, direct minimization of the loss function in
 \ref{objectiveFunction}.)  Since our model is derived from the same
 loss function it is expected that the two methods would give similar
 estimates, and this is confirmed by our simulations. Indeed, the
 parameter estimates from both methods are nearly identical (the small
 differences are attributed to the chosen tolerance level of the
 computational methods and the fact that empirical quantiles are not
 uniquely defined). \revise{For example, in scenario \#13 there are five predictors,
 and for each quantile $q\in\{0.05,0.1,\ldots,0.95\}$,
 the absolute mean difference between the two estimators  across all five predictors is 0.0003 (using 100 
 simulated datasets). Similar results are observed in all scenarios.
 }
 
 The variances of the regression parameter
 estimates from the two methods, however, are quite different.  Recall
 that we obtain the asymptotic covariance of $\hat{\bm\beta}$ by using
 Bahadur's representation, which requires the estimation the density
 of the residual $u_i$ at 0. To do that, we use kernel density
 estimation, as implemented in the \texttt{KernSmooth}
 package \citep{KernSmooth}.  See \cite{Deng2014DensityEI} for
 a review of kernel density estimation packages.  The
 \texttt{quantreg} package computes confidence intervals using the
 inversion of a rank test, per
 \citep{koenker1994confidence}.  Figure \ref{sdbeta1} shows
 $\hat\sigma_{\hat\beta_1}$ for $q\in\{0.05,0.1,\ldots,0.95\}$, using
 QREM (blue squares, left), the bootstrap (dark-red triangle, middle),
 and \texttt{quantreg} (orange circles, right). The horizontal yellow
 lines above each quantile represent the true estimate of the standard
 deviation of $\hat\beta_1$, as obtained from the asymptotic
 Bahadur-type estimation using the true density of $u_i$ at 0.  Our
 estimator has a smaller sampling variance across all scenarios and
 all quantiles.  Figures \ref{sdbeta1sim15} and \ref{sdbeta1sim21} in
 Appendix \ref{sec.appendix} show the estimated standard deviations of
 $\hat\beta_1$ obtained from QREM and \texttt{quantreg} for scenarios
 15 and 21, respectively.  Again, the sampling variance of
 $\hat\sigma_{\hat\beta_1}$ obtained from QREM is smaller than the one
 from \texttt{quantreg}, and especially near the edges ($q<0.15$ and
 $q>0.85$.)  These results show that while on average the coverage
 probability of the two methods is expected to be similar, the smaller
 variance obtained from the QREM procedure imply that results from
 this method are more stable and provide more reliable inference. The
 wider range of variance estimates obtained from \texttt{quantreg}
 imply that this method is much more likely to either lack power or to be
 over-powered.

\begin{figure}[t!]
  \begin{center}
 \includegraphics[trim={2cm 5.5cm 6cm 1.5cm},clip,scale=0.6]{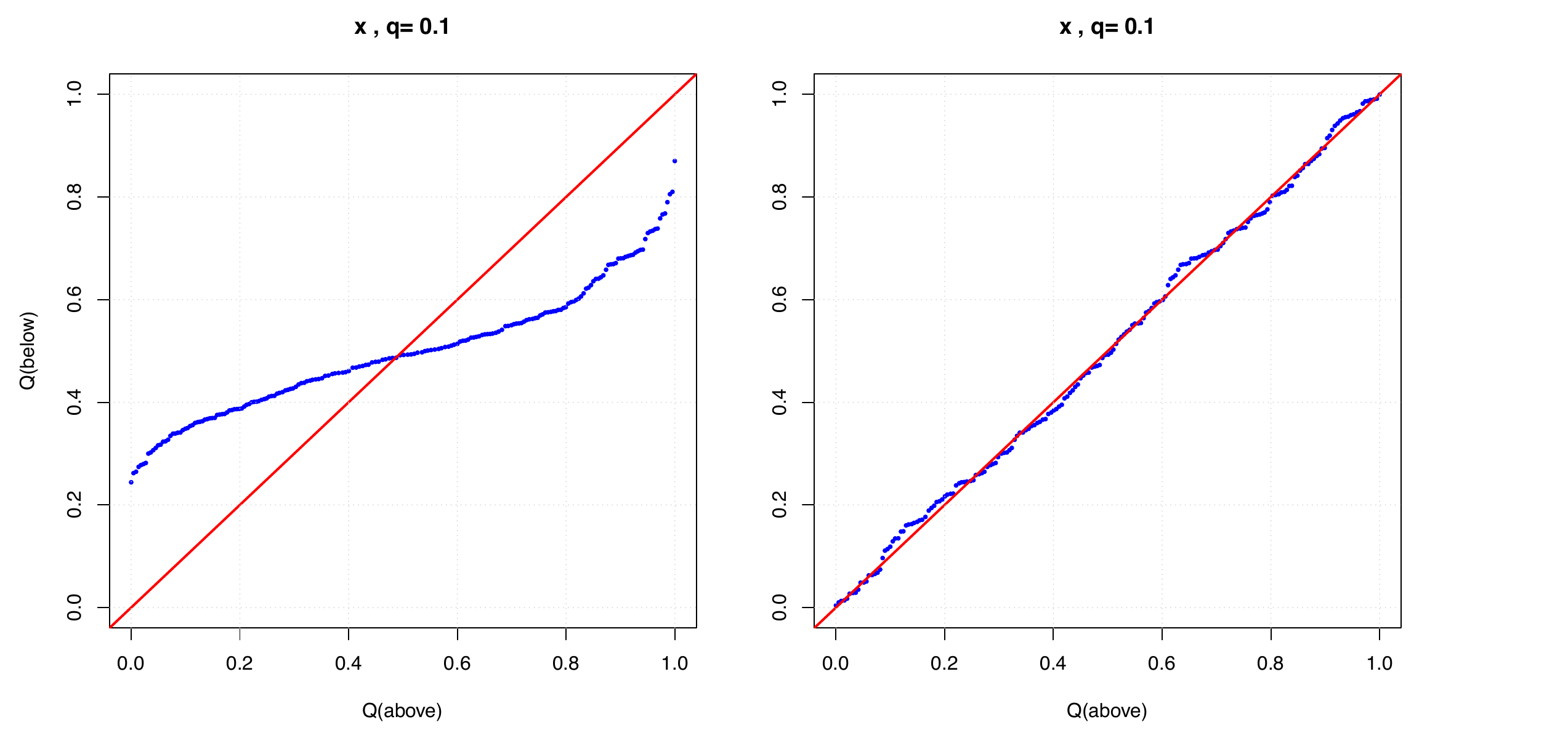}
\end{center}
\caption{QR simulation - the true model is $y \sim N(6x^2+x+120, (0.2+x)^2)$. Showing
quantile-quantile plots for the linear predictor $x$ when fitting a linear  (left) and the correct, quadratic model (right),
for $q=0.1$.}\label{quadraticQQQ}
 \end{figure}

 We also simulate data in which the response depends on predictors in
 non-linear ways.  For example, in Scenario \#23, 
 $y \sim N(6x^2+x+120, (0.2+x)^2)$ so that the relation between the
 mean of $y$ and $x$ is a quadratic function, and the standard
 deviation increases linearly with $x$.  To assess model adequacy, we
 use the quantile-quantile plot construction as described in Section \ref{sec.model}
 and the goodness of fit definition in equation
 (\ref{GOF}).  For example, Figure \ref{quadraticQQQ} shows
 the Q-Q plot for the predictor $x$ when $q=0.1$: on the left hand
 side, we fit a linear model, $y\sim x$, and on the right hand side
 the fitted model is $y\sim x + x^2$ (the true model). The Q-Q plot
 suggests that, for the 10th percentile, the linear model is
 inadequate, while the quadratic one fits very well. The goodness of
 fit, as defined in (\ref{GOF}), is $G=1,135$ for the
 linear model, and $G=828$ for the quadratic model, again providing
 evidence that in this case a quadratic model provides a better fit.
  
\begin{figure}[ht!]
 \advance\leftskip-2cm
  \includegraphics[width=1.25\textwidth]{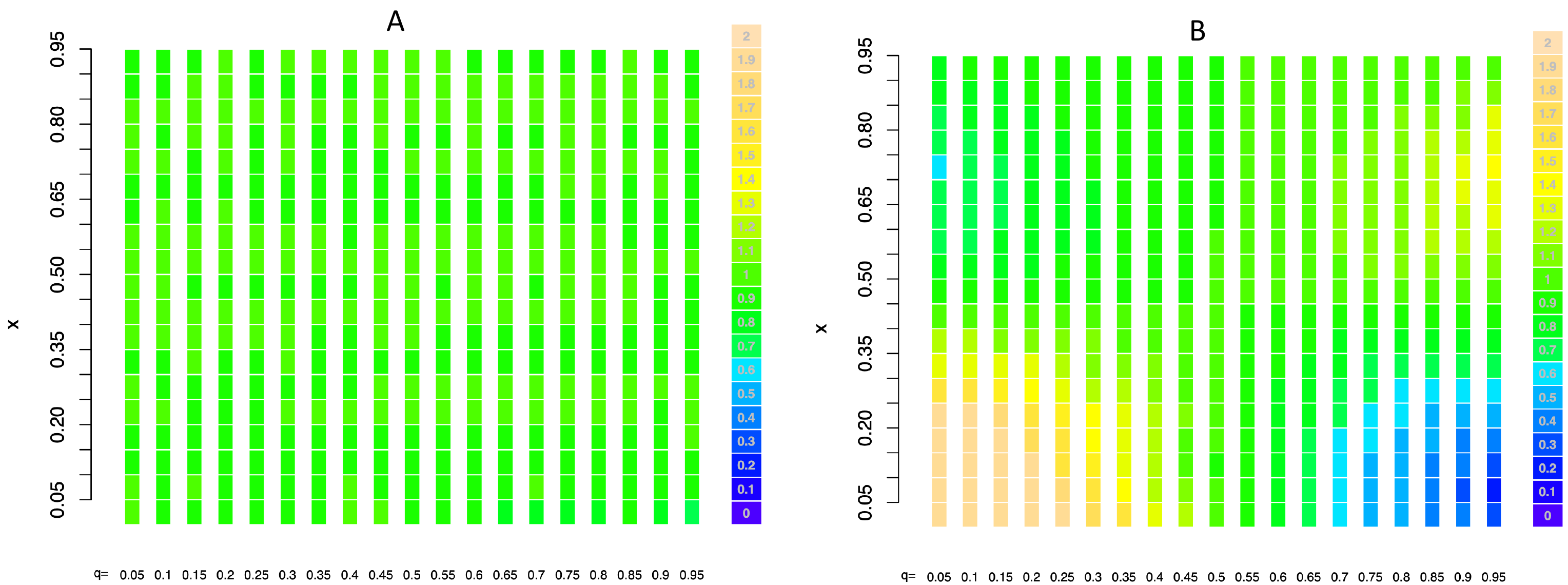}
 \advance\rightskip1cm
\caption{A `flat' Q-Q  plot for $q\in(0.05,0.1,\ldots,0.9,0.95)$, when the true model is
$y\sim N(4x_1x_2, (0.1+0.2x_1)^2)$. A: fitting the correct (interaction) model. B. Fitting an additive (incorrect)
model, $y\sim x_1+x_2$.}\label{QxQy}
 \end{figure}

 It is possible that a model would fit well for certain quantiles, but
 not for others. The next example demonstrates this point and shows an
 effective way to visualize multiple Q-Q plots, when fitting the same
 quantile regression model for different $q$'s.  We simulate
 10,000 points from a bivariate uniform distribution on
 $[0,1]\times[0,1]$ as our two predictors, $x_1$ and $x_2$, and
 generate the response using the interaction of the predictors, so
 that $y\sim N(4x_1x_2, (0.1+0.2x_1)^2)$ (simulation 24 in the
 Appendix).  For each $q\in(0.05,0.1,\ldots,0.9,0.95)$ we fit two
 models - one additive, $y\sim x_1+x_2$, and one with an interaction
 term, $y\sim x_1+x_2+x_1x_2$.  From each fitted model we obtain the
 theoretical and empirical quantiles, $Q_{q,t}(x_1)$, and
 $Q_{q,e}(x_1)$, respectively, with respect to $x_1$.  Recall that
 under the correct model the plot of the theoretical versus empirical
 quantiles should lie close to the $45^{\circ}$ line. So, for a fixed
 $q$, and some $\xi_1$ in the range of $x_1$ we define
 $r_q(\xi_1)=n_{q,e}(\xi_1)/n_{q,t}(\xi_1)$ where $n_{q,e}$ and
 $n_{q,t}$ are the numbers of empirical and theoretical quantiles that
 are smaller than $\xi_1$.  An adequate model gives
 $r_q(\xi_{1})\approx 1$ for each value of $\xi_1$.  For each $q$ we
 use $L$ (e.g., 20) equally spaced values in the range of $x_1$,
 denoted by $\xi_{1j}$, and obtain $r_q(\xi_{1j})$.  We plot an array
 of rectangles with colors corresponding to the values of
 $r_q(\xi_{1j})$, so that the columns in the array correspond to the
 quantiles, and the rows to $\xi_{1j}$.  This yields a heatmap, as
 depicted in Figure \ref{QxQy}, to which refer as a `flat' Q-Q plot,
 since for each $q$ we convert the two-dimensional Q-Q plot to a single
 column in the heatmap.  Figure \ref{QxQy}-A shows an ideal `flat'
 Q-Q plot. The interaction model fits very well for each $q$.  In
 contrast, Figure \ref{QxQy}-B shows that although the additive model
 suggests a good fit for values around $q=0.5$, it is inadequate for
 most $q$'s.

 We also simulate data from mixed models. In simulation 25 we
 generate a random sample of $100$ independent subjects, and each
 subject was observed at four time points. Within subject the
 observations are correlated.  That is, we use a variance component
 model, where $y_{it}=2 + x_{it} + u_i + \epsilon_{it}$, and the
 random effect, $u_i$, is distributed as $N(0,0.5^2)$, and the random
 errors as $ \epsilon_{it}\sim N(0, 0.1^2)$.  To obtain confidence
 intervals for the parameter estimates we use a bootstrap approach
 and fit the QR model 1,000 times for each $q=\{0.1,\ldots,0.9\}$,
 each time drawing a random sample (with replacement) from the
 subjects.  The parameter estimates and the bootstrap standard errors
 are very accurate for all deciles: the average bias for $\beta_1$
 is 0.0008, and the 95\% coverage probability is $0.95\pm 0.005$.

 To assess the performance of Algorithm \ref{algorithmLargeP} in the
 large $P$ setting, in each scenario in Table
 \ref{sec.appendix}\ref{simsLargeP} we use $n=200$ subjects and the
 true predictors are drawn from a standard uniform distribution.
 Each data set is augmented to include 500 predictors, such that the
 ones not related to $Y$ are drawn as i.i.d $N(0,0.1^2)$. Each
 configuration is simulated 100 times.  We then run Algorithm
 \ref{algorithmLargeP} with $q\in\left\{0.1,0.2,\ldots,0.9\right\}$
 and evaluate the performance of our method in terms of the total
 number of true/false positive/negative.  We also run a lasso-based
 quantile regression variable selection method as implemented in the recent
 \texttt{rqPen} package \citep{rqPen}. We set the tuning parameter
 $\lambda=0.1$, because it appeared to yield relatively good results
 in the different scenarios (for example, with $\lambda=0.01$ we had
 too many false positives, and with $\lambda=1$ no predictors were
 selected. The automated approach of finding $\lambda$ via cross
 validation implemented in the \texttt{cv.rq.pen} function, proved to
 be much too time-consuming\footnote{For example, to complete one run
   of cv.rq.pen with $P=500$ and $n=200$ and the default function
   setting took 33 minutes on a Linux OS with eight i7-4710MQ CPUs @
   2.5GHz, four cores. With 100 replications for each combination of
   scenario and quantile, performing cross validation for each run was
   impractical.}  with a large $P$.)

 In Table \ref{sim5} we show results for Simulation \#5 from the list
 of scenarios listed in Table \ref{sec.appendix}\ref{simsLargeP}.  The
 results represent the other simulation cases. In this scenario the
 standard deviation increases linearly with one of the predictors, and
 the number of true predictors is five.  Using our approach, at least
 two true predictors are found in all the simulations and except for a
 small number of cases with $q=0.1$ and 0.9, at least four true
 predictors are found.  For $q\in[0.3,0.7]$ all five predictors are
 found at least 93\% of the time. With the lasso-based method no
 predictors are found 38\% of the time when $q=0.1$ and 73\% when
 $q=0.9$. In general, in all the simulations the true-positive rate
 when $q<0.25$ and $q>0.75$ is lower when using the lasso than with
 our approach.

 In terms of the false positive rate, it can be seen in the lower
 section of Table \ref{sim5} that with SEMMS, in the vast majority of
 cases there are no false positives and only in very few cases when
 $q=0.1, 0.3,$ or 0.9 are there two or three false positives. In
 contrast, falsely detected predictors are more common with the
 lasso.  For example, when $q=0.5$, the lasso-based method yields
 three or more false positives 35\% of the time.  While it had no
 false positives for $q=0.1$ and 0.9, it also has very low power for
 these quantiles.

\begin{table}[b!]
\centering
\resizebox{\columnwidth}{!}{
\begin{tabular}{|r|rrrrrrrrr||rrrrrrrrr|}
   \hline
TP   & \multicolumn{9}{c||}{SEMMS} & \multicolumn{9}{c|}{LASSO}\\
       & 0.1 & 0.2 & 0.3 & 0.4 & 0.5 & 0.6 & 0.7 & 0.8 & 0.9 & 0.1 & 0.2 & 0.3 & 0.4 & 0.5 & 0.6 & 0.7 & 0.8 & 0.9 \\ 
  \hline
  5 &  53 &  66 &  93 &  96 &  96 &  98 &  96 &  49 &  24 &     &  10 &  73 &  95 &  98 & 100 &  92 &  40 &   \\ 
  4 &  44 &  34 &   7 &   4 &   4 &   2 &   4 &  51 &  70 &     &  71 &  27 &   5 &   2 &     &   8 &  57 &   \\ 
  3 &   2 &     &     &     &     &     &     &     &   6 &     &  13 &     &     &     &     &     &   3 &    \\ 
  2 &   1 &     &     &     &     &     &     &     &      &     &   6 &     &     &     &     &     &     &   3 \\ 
  1 &      &     &     &     &     &     &     &     &      &      62 &     &     &     &     &     &     &     &  24 \\ 
  0 &      &     &     &     &     &     &     &     &      &      38 &     &     &     &     &     &     &     &  73 \\ 
   \hline
   \hline
FP   & \multicolumn{9}{c||}{SEMMS} & \multicolumn{9}{c|}{LASSO}\\
       & 0.1 & 0.2 & 0.3 & 0.4 & 0.5 & 0.6 & 0.7 & 0.8 & 0.9 & 0.1 & 0.2 & 0.3 & 0.4 & 0.5 & 0.6 & 0.7 & 0.8 & 0.9 \\ 
  \hline
  0 &  62 &  95 &  95 & 100 &  99 & 100 &  99 & 100 &  93   & 100 &  90 &  39 &  20 &  20 &  17 &  50 &  83 & 100 \\ 
  1 &  32 &   5  &    4 &        &    1 &        &    1 &        &   6    &        &    9 &  47 &  38 &  24 &  31 &  32 &  15 &  \\ 
  2 &   4  &       &    1 &        &       &        &       &        &   1    &        &    1 &  10 &  23 &  21 &  24 &  12 &    1 &\\ 
  3 &   2  &       &       &        &       &        &       &        &         &        &       &    4 &  11 &  20 &  20 &    3 &    1 &\\ 
  4 &       &       &       &        &       &        &       &        &         &        &       &       &    6 &    9 &    7 &    2 &       & \\
  5 &       &       &       &        &       &        &       &        &         &        &       &       &    1 &    5 &    1 &    1 &       & \\
  6 &       &       &       &        &       &        &       &        &         &        &       &       &       &    1 &       &       &       &\\
  7 &       &       &       &        &       &        &       &        &         &        &       &       &    1 &       &       &       &       &\\
   \hline
\end{tabular}
}
\caption{Simulation \#5 -- the true number of predictors is 5 out of P=500, sample size N=200, number of simulations B=100.
Top: number of simulations in which $k=0,\ldots,5$ true predictors were found by SEMMS (left) and the lasso (right)
for $q=0.1,\ldots,0.9$. Bottom: number of simulations in which $k=0,\ldots,7$ false predictors were found by SEMMS and the lasso.}\label{sim5}  
\end{table}

We also show results for simulation \#9 from the scenarios listed in
Table \ref{sec.appendix}\ref{simsLargeP}. Here, we use a more
challenging setting with $P=1,000$ and $n=100$. In this scenario
$Y=X_1+\ldots+X_{20}+\epsilon$ where $\epsilon\sim N(0,0.1^2)$, but,
unlike the previous example the twenty predictors are all correlated
with an autoregressive structure (AR1) with $\rho=0.95$. The results
are summarized in Table \ref{sim9}, where we show the median number of
correctly and incorrectly detected predictors for each decile, for our
approach (left) which uses the SEMMS variable selection algorithm, and
the lasso-based approach (right) as implemented in the \texttt{rqPen}
package using $\lambda=0.1$.
Our approach is more powerful and yields almost no false
positive predictors, while the median number of false positives with
the lasso is 20 when $q=0.5$.

\begin{table}[ht]
\centering
\resizebox{\columnwidth}{!}{
\begin{tabular}{|r|rrrrrrrrr||rrrrrrrrr|}
  \hline
     & \multicolumn{9}{c||}{SEMMS} & \multicolumn{9}{c|}{LASSO}\\
     & 0.1 & 0.2 & 0.3 & 0.4 & 0.5 & 0.6 & 0.7 & 0.8 & 0.9  & 0.1 & 0.2 & 0.3 & 0.4 & 0.5 & 0.6 & 0.7 & 0.8 & 0.9 \\ 
  \hline
  TP & 20 & 20 & 20 & 20 & 20 & 20 & 20 & 20 & 20   & 12 & 15 & 15 & 16 & 16 & 16 & 15 & 14 & 11 \\ 
  \hline
  FP & 0 & 1 & 0 & 0 & 0 & 0 & 0 & 1 & 0   &  0 & 4 & 12 & 18 & 20 & 18 & 12 & 4 & 0 \\ 
   \hline
\end{tabular}}\caption{Simulation \#9 -- the true number of predictors is 20 out of P=1,000, sample size N=100, number of simulations B=100. The true predictors have an AR(1) correlation structure with $\rho=0.95$.
The median number of true and false predictors found by SEMMS (left) and the lasso with $\lambda=0.1$
(right) for $q=0.1,\ldots,0.9$.}\label{sim9}
\end{table}

\section{Case Studies}\label{sec.casestudy}
\subsection{Frailty QR Model -- Emergency Department Data}
\label{long_data}
The National Center for Health Statistics, Centers for Disease Control and Prevention
conducts annual surveys to measure nationwide health care utilization.
We use the 2006 NHAMCS (National Hospital Ambulatory Medical Care Survey) data to demonstrate
fitting quantile regression models with random effects to survival-type data.
In this case, the length of visit (LOV, given in minutes) while in a hospital emergency department (ED)
is considered the `survival' time, and we define the normalized response $y=\log_{60}(LOV+1)$.
We filter the data and remove hospitals with fewer than 70 visits, which results in 21,262 ED visit records from 230
hospitals.

First, we fit a fixed effect model which includes nine predictors of interest:
Sex, Race (white, black, other), Age (standardized to a $[0,1]$ range), Region (northeast, midwest,
south, and west), Metropolitan area (yes/no), Payment Type (Private, Government/Employer, Self, and Other),
Arrival Time (8AM-8PM or 8PM-8AM), the Day of the Week, and a binary variable (Recent Visit) to indicate
whether the patient has been discharged from a hospital in the last 7 days, or from an ED within the last 72 hours.
We fit a QR (fixed effect) model for $q=0.025, 0.05, 0.075,\ldots,0.975$.
Then, we fit the frailty model for the same quantiles using the same nine predictors, plus
the Hospital as a random effect, and check whether the coefficients of the fixed effects change
when we account for the correlation between patients within a hospital.

\begin{figure}[t!]
  \begin{center}
 \includegraphics[trim={1cm 6cm 4cm 3.5cm},clip,width=0.9\textwidth]{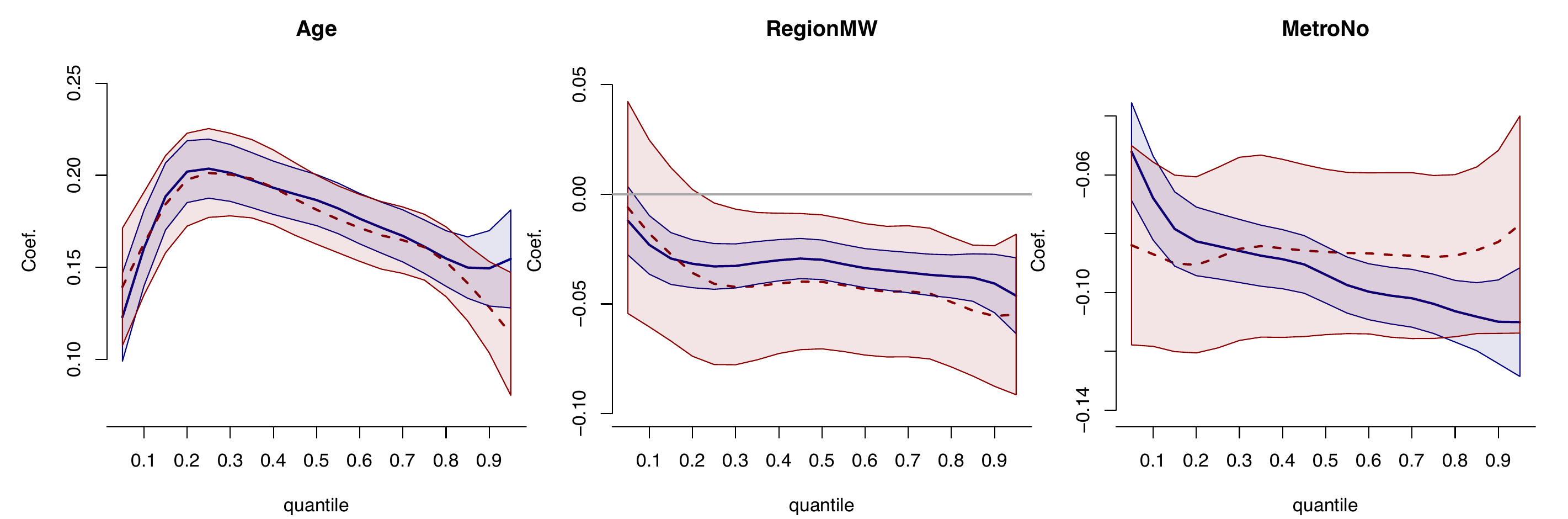}
\end{center}
\caption{ED data -- quantile regression coefficients with 95\% point-wise confidence intervals, 
from left to right, for Age, Region (=midwest, where northeast is the
 baseline), and Metro (=No, where Yes is the baseline). The blue, solid line is the for the fixed effect model
 coefficients, and the red, dashed line is for the mixed model coefficients.}\label{ERf3}
 \end{figure}
 
 Figure \ref{ERf3} shows the regression coefficients, from left to right, for Age, Region (=midwest, where northeast is the
 baseline), and Metro (=No, where Yes is the baseline). For these variables, the coefficients remain approximately the
 same when we add the Hospital random effect. Age has a positive effect for all quantiles, and the smallest difference in LOV
 due to age is for the patients who are discharged quickly from the ED, or those who stay at the ED the longest.
 Generally, ED patients in the northeast have longer visits than the ones in the midwest (except for those
 who are discharged quickly). Similarly, ED patients in a metropolitan area stay significantly longer than in
 rural area hospitals.
 The Arrival Time and Day of Week variables are also similar whether we include the hospital random effect or not (not
 shown here.) Arriving at an ED on Monday results in a longer stay compared with Sunday (except for patients who 
 are discharged quickly), but the difference between other weekdays and the weekend is less significant for most
 quantiles. Arriving at night means a longer stay only for the patients who end up staying the longest (the upper 
 20-th percentile in the fixed effect model but no significant difference in the mixed model), but among patients
 who are discharged relatively quickly ($q\le0.2$) arriving at night actually corresponds to a shorter stay, as compared
 with day-time arrival.

\begin{figure}[ht!]
  \begin{center}
 \includegraphics[trim={3cm 6cm 4cm 3.5cm},clip,width=0.9\textwidth]{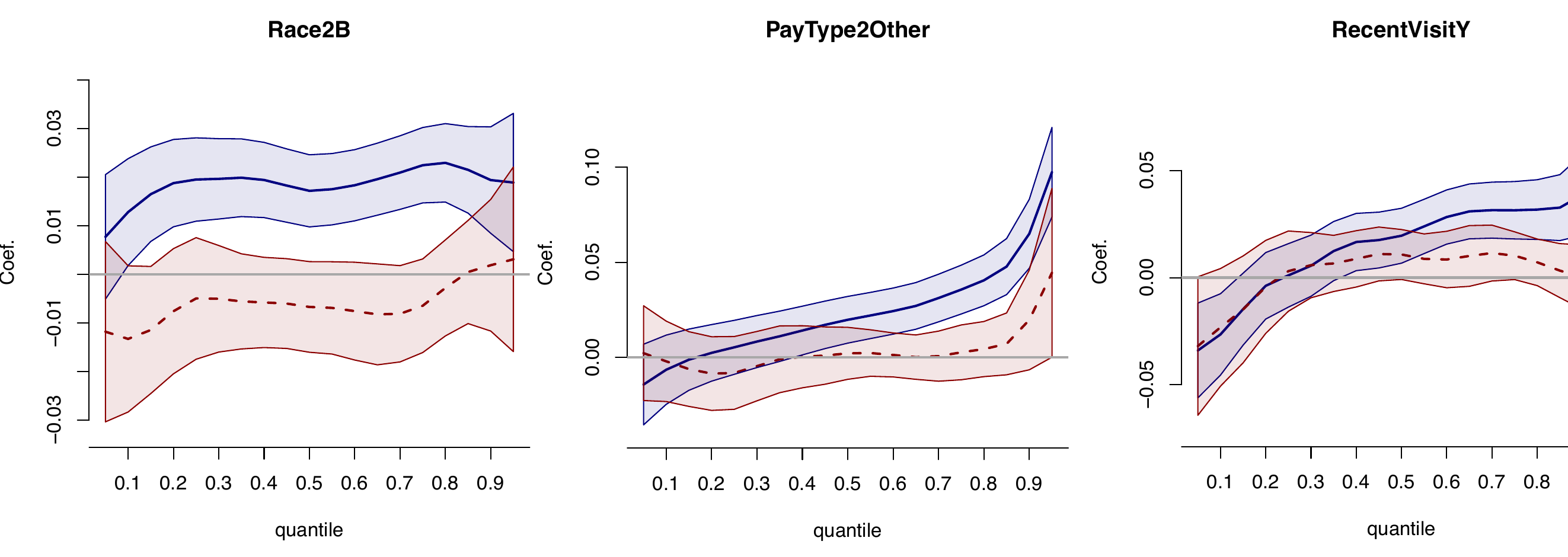}
\end{center}
\caption{ED data -- quantile regression coefficients with 95\% point-wise confidence intervals,
for Race=Black (left, with White as baseline),  Pay Type (center, for `Other' source of payment versus `Private'),
and Recent Visit (for Yes, versus the No reference.)
The blue, solid line is the for the fixed effect model
 coefficients, and the red, dashed line is for the mixed model coefficients.}\label{ERf2}
 \end{figure}

Figure \ref{ERf2} shows that accounting for within-hospital correlation yields
very different results compared with the fixed effect model. According to the fixed effect model
one might conclude that there is a significant difference between blacks and whites (left panel)
and between people with private health insurance versus people whose medical bill is `Other' (namely, 
not Private, Medicare, Medicaid/SCHIP, Worker's Compensation, or Self pay), for patients who are 
not discharged quickly, i.e., for $q>0.2$. Similarly, from the fixed effect model one might conclude that among
those who are not discharged quickly, a recent visit to a hospital or an ED will result in longer stays (for $q>0.35$).
However, the results from the mixed model suggest that these differences can be explained by variation between 
hospitals, while, within a hospital there is no significant difference in the length of visit
between black and white people, or between people with private insurance and people who `other' pay.

 \subsection{The TCGA Data -- Lung Cancer}
We demonstrate  our approach to quantile regression in the `large P' setting with the Cancer Genome Atlas (TCGA) dataset. We use a subset
of 379 lung cancer patients who were either lifetime smokers or have been reformed smokers for less than 15
years. 
We excluded non-smokers in order to work with a more homogenous group, in terms of possible
genetic damage. We excluded genes with zero expression in at least one sample, and the number
of genes remaining in the analysis was 13,492.
\begin{figure}[b!]
\begin{center}
\includegraphics[trim={0cm 0cm 0cm 2cm}, clip, width=8cm]{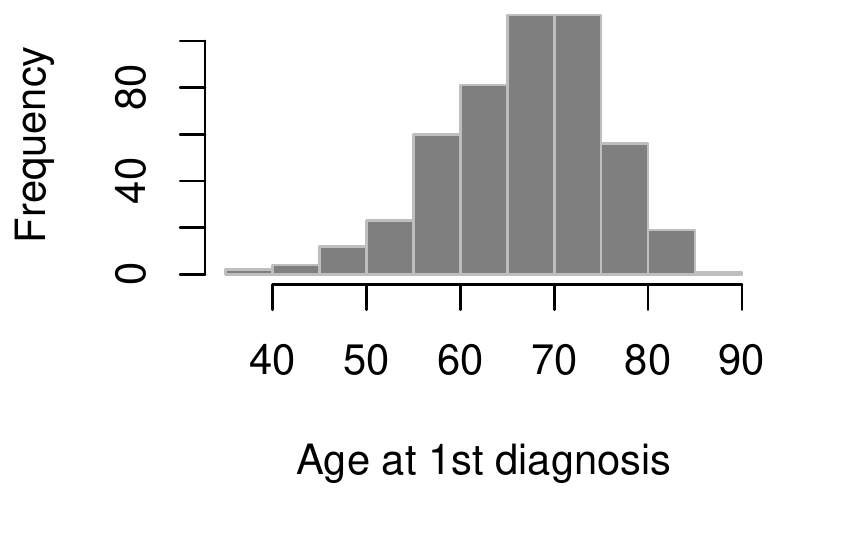}
\caption{The distribution of the age at first diagnosis of 379 lung cancer patients from the TCGA dataset.}
\label{fig:agedx}
\end{center}
\end{figure}
Figure \ref{fig:agedx} shows the distribution of the age at first diagnosis of the 379 smokers. It can
be seen that the ages range from 39 to 90. We wanted to check whether any genes are 
associated with the onset of lung cancer (age at first diagnosis). In particular, genes associated
with an early onset  may be good candidates for being targeted, and  genes
associated with a late onset of lung cancer despite the smoking habits may help shed light on the 
potentially important protective pathways.
We performed variable selection for three quantiles: $q=0.15$, 0.5, and 0.85.
We do not find any predictors for the median age, $q=0.5$, but among the 13,492 genes we
find one significant predictor for the 15th, and one for the 85th quantile.

We identify SCO1 (Entrez 6341) as the only predictor for $q=0.15$, and obtain the following relationship
between the log-expression of SCO1 and age at first diagnosis:
$$Q_{0.15}(age)=58.1 - 2.27\cdot\log_2(SCO1).$$
The standard error for the effect of the SCO1 gene is 0.616 (note that this estimate does not account for selection bias).  This finding is quite reasonable since the SCO1 gene plays a critical role between cooper (Cu) incorporation into the cytochrome c oxidase (CCO) and lung cancer.  \cite{suzuki2003identification} report evidence that the CCO assembly protein COX17 is a potential molecular target for treatment of lung cancers. COX17 and SCO1 are involved in Cu incorporation into the  CCO.  Partially oxidized COX17 in the intermembrane space hands off Cu and two electrons to oxidized SCO1 \citep{banci2008mitochondrial}. Cells deficient in SCO1 have cytoplasmic Cu deficiency but normal mitochondrial Cu content, suggesting the existence of homeostatic mechanisms governing Cu distribution \citep{dodani2011targetable}.


For $q=0.85$ we identify the gene PTGES3 (Entrez 10728) with the following relationship with age,
with a standard error of 0.43 for the coefficient of the gene:
$$Q_{0.85}(age)=74.1 - 1.26\cdot\log_2(PTGES3).$$
Prostaglandin E synthase 3 is an enzyme that in humans is encoded by the PTGES3 gene. The protein encoded by this gene is also known as p23 which functions as a chaperone which is required for proper functioning of the glucocorticoid and other steroid receptors \citep{freeman2002disassembly}. The heat shock protein 90 (Hsp90) has a critical role in oncogenic survival signaling. Overexpression of Hsp90 in human cancer cells correlates with poor prognosis \citep{zhang2004targeting}. The most important interactions within the Hsp90 chaperone system are between Hsp90 isoforms and co-chaperone p23, which occur only when Hsp90 is bound to ATP. These interactions are crucial for maturation of Hsp90 protein complexes and the release of folded proteins \citep{myung2004expressional}.


\subsection{The Riboflavin Data}\label{casestudy:ribo}
We demonstrate a graphical model application of our method, using the riboflavin data made available 
and analyzed by \cite{buhlmann2014}.
The dataset contains 71 samples with (normalized) expression data for 4,088 genes. 
The response variable is the riboflavin production rate in \textit{Bacilluss subtilis}. Our goal is to construct 
a graphical model for different quantiles of the response. Since $n$ is rather small, we use only $q=0.25$, 0.5, and 0.75.
First, for each of the quartiles we use Algorithm \ref{algorithmLargeP} to perform variable selection for the quantile regression model 
with riboflavin as the response and all 4,088 genes as putative predictors. Then, for each
gene, we run Algorithm \ref{algorithmLargeP} with that gene as the response and the other 4,087 genes as putative predictors.
We present the resulting graphical model for the 25th percentile of vitamin $B_2$ production rate.

All the fitted models are sparse, with at most nine predictors. Only a small number of genes are not found to be associated with
any other gene (5, 10 and 2 for $q=0.25$, 0.5, and 0.75, respectively.) Hence, the network is sparse but highly interconnected,
which suggests that the relationship between the response and the genes may be quite complex and probably cannot
be adequately explained by a simple additive regression model.
To keep the interpretation of the graphical model as simple as possible, we present the subnetwork around the response, and limit
the path length away from the center to three edges. 
The results are shown in Figure \ref{fig:graphicalmodel}. There are four genes associated with the 25th percentile of riboflavin
production rate directly (XHLB, YCKE, ILVD, and YXLD). These effects are shown as dashed lines.
A solid edge in the graph represents  a significant relationship between genes for at least one of the quartiles. If a relationship
between a pair of genes is found to be significant in all three quartiles, no label is added to the 
corresponding solid line. Otherwise,
the quartiles for which the relationship is found to be significant appear along the solid edge.
A red, thick edge represents a negative effect, while a blue, thin edge represents a positive one. 
Some genes are connected via a bi-directional edge, indicating that both were found to be a strong predictor of the other.
For example, YXLD has a negative effect on $B_2$ and its expression is related to that of YXLG in all three quartiles.
LYSA has a positive effect on ILVD, but only for $q=0.25$ and 0.5, and ILVD is not associated with the expression of LYSA.  The orange dots near a labeled gene in Figure \ref{fig:graphicalmodel} represent genes that are highly correlated with the listed gene.  The genes represented by orange dots are often in the same regulatory pathway as the listed gene.

We construct similar graphs for the median and the third quartile of $B_2$ (not shown here) and observe that the 
XHLB and MTA branches are common to all three quartiles, and although the YXLD is not, a closely related pair of
 genes (namely, YXLE and YXLF) replace that branch for $q=0.5$ and 0.75.
The YQKJ branch, however, is unique to the lower quartile of vitamin $B_2$ production rate. Similarly, 
the third quartile has a unique branch with YFKN as a strong predictor for $Q_{0.75}(B_2)$.
Obtaining these paths which are unique to very low or very high levels may help understand how 
to regulate the $B_2$ production rate. 

\input{"drawtikz"}
We use the code in  \cite{buhlmann2014} to reproduce their network diagram in Section 4.3,
where they restrict the data set to a set of 100 genes with the largest empirical variance. Using the `huge' package
 \cite{huge} we obtain the following neighbors of $B_2$, when considering the mean as the response:
YXLE,  XHLA, XHLB, YCKE, YTGD, YHZA, YCGN. The first four are found in our network. However,
when we repeat this analysis with all 4,088 genes and the huge package, no edges are found in the network.  Thus, our method not only allows to construct graphical models for specific quantiles, but it also appears to  yield more discoveries than with the huge package when $P$ is large.

The \textit{SubtiWiki} at \url{http://www.subtiwiki.uni-goettingen.de/} gives a detailed functional network annotation of genes and proteins in \textit{Bacillus subtilis}.  \textit{SubtiWiki} is based on a relational database and provides access to published information about the genes and proteins of \textit{Bacillus subtilis} and about metabolic and regulatory pathways.  The graphical model in Figure \ref{fig:graphicalmodel} corresponds to the underlying functional network annotations.   RpoA, RpoB, and RpoC are related to transcription and the sigma factors are the modules that do the work. YXLG and YXLD (along with the neighboring orange dots YXLC, YXLE, YXLF, and YBGB)
are in the sigY regulon and XHLB and XHLA (along with the neighboring orange dots XTMA, XTMB, XKDE, XKDF, XKDG, XKDH, XKDI, XKDJ, XKDK, XKDM, XKDN, XKZB, XKDO, XKDP, XKDQ, XKDR, XKDS, XKDT, XKDU, XKZA, XKDV, XKDW, XKDX, XEPA, and XLYA) 
 are in the XPF regulon.  For a graphical representation of the sigY and XPF regulons, see, respectively: 
\sloppy\url{http://www.subtiwiki.uni-goettingen.de/v3/regulation/view/801E92306971E26AD4AB155172B7F4EFDE2F9170}
and \url{http://www.subtiwiki.uni-goettingen.de/v3/regulation/view/458406A4E6824C24493CFC19718F10720AA3B453}.

 Many of the other genes are in the sigA, codY or Spo0A regulons. These sigma factors fit together in networks for transcription. It turns out that YQKJ and YCKE are also known as MLEA and BGLPH, respectively.  MLEA is connected to the CCPA regulon and BGLPH is related to CCPA repression.  The negative sign on YQKJ may be due to the YCKE repression effect.  Also ILVD is in the codY regulon and YQKJ is in the CCPA and they interact along with RpoA.  ILVD is a very connected gene and has a lot of neighbors in the quartile regression network. This branch is unique to the lower quartile, and it looks like the big network around ILVD interacts with riboflavin via YQKJ. As the expression in the ILVD network increases,  YQKJ increases too, and that causes the lower quartile of riboflavin to decrease (but not the median or the upper quartile.)

\section{Discussion and Future Work}\label{sec.discussion}

The two main contributions in this article are the development of a flexible mixed effect modeling approach and the variable selection tools for QR.  We exploited the fact that the estimating equations for QR can be 
solved using a simple EM-type algorithm in which the M-step is computed via weighted least squares, with weights computed at the E-step as the expectation of independent generalized inverse-Gaussian variables.
Because the M-step involves fitting an ordinary linear model it is  straightforward to extend the algorithm to allow for random effects in the linear predictor.  The computational approach is compared with existing software using simulated data, and the methodology is illustrated with several case studies. 

In Section \ref{long_data} we highlighted the utility of the mixture representation approach to longitudinal data case study. However, for data collected on dense grids, traditional longitudinal regression approaches are not directly applicable since the number of grid points may be larger than the sample size and the correlation between the dependent variable on the distinct grid points may be quite high. 
Estimation methods such as smoothing splines and a reproducing kernel Hilbert space approach could also be adapted for functional quantile regression and the mixture representation approach could likely be applied here too.  In Gaussian functional data settings \cite{ramsay1997functional} use a penalized regression which is closely connected to a random effects model.  We conjecture that a variant of Algorithm \ref{algorithm2} can be applied to directly solve the functional quantile regression estimating equation.

In Section \ref{casestudy:ribo} we constructed a graphical model to elucidate the structural inter-dependencies in the riboflavin gene expression network.  It was critical in the model selection procedure to use different quantiles (q=0.25, 0.5, 0.75) rather than a single fixed level.  There has been a lot of recent interest in multiple quantile graphical model \citep{ali2016multiple, belloni2011, belloni2016quantile, karpman2018learning}.  These models are essentially fitted using a quantile version of the neighborhood selection approach of \cite{meinshausen2006} for learning sparse graphical models, which is equivalent to variable selection for penalized quantile regression models. Algorithm \ref{algorithmLargeP} gives a direct and scalable approach to edge selection for multiple quantile graphical models via the tractable mixture representation.

\cite{chen2018quantile} recently developed quantile factor models. Unlike traditional factor models that represent the latent structure as mean-shifting factors, the new approach recovers unobserved factors by shifting other relevant parts of the distributions of observed variables. However, the computational algorithm involves iterative quantile regressions using linear programming methods. The mixture representation approach could likely be applied as computational and inferential approaches to the quantile factor models using a modification of Algorithm \ref{algorithm} via an additional layer of an EM algorithm to estimate a Gaussian factor analysis with regression analysis \citep{rubin1982algorithms}.

\bibliographystyle{apalike}

\bibliography{qrem}
\clearpage
\pagebreak

\appendix
\section{Appendix}\label{sec.appendix}

\subsection{The binary-valued QR residuals satisfy $\bX^T\hat\bc\equiv 0$}\label{propUncorr}
We denoted  the scaled, binary-valued QR
residuals by $\mathbf{c}=sgn(Y-X \bm\beta_q) - (1-2q)\mathbf{1}$. We
found the WLS
solution from the M-step:
$$\bm\beta_q = (X^T\Lambda^{-1}X)^{-1}X^T\Lambda^{-1}\left[Y-(1-2q)\bm\lambda\right].$$
Multiplying both sides by $(X^T\Lambda^{-1}X)$ we get
$$(X^T\Lambda^{-1}X)\bm\beta_q = X^T\Lambda^{-1}\left[Y-(1-2q)\bm\lambda\right]$$
and rearranging terms we get
\begin{eqnarray*}
X^T\Lambda^{-1}(Y-X^T\bm\beta_q)&=&(1-2q)X^T\Lambda^{-1}\bm\lambda\\
&=&(1-2q)X^T\mathbf{1}\,.
\end{eqnarray*}
Expressing $Y-X \bm\beta_q$ as $\Lambda\cdot sgn(Y-X \bm\beta_q)$ we get
\begin{eqnarray*}
X^T\Lambda^{-1}[\Lambda\cdot sgn(Y-X\bm\beta_q)]&=&(1-2q)X^T\mathbf{1}
\end{eqnarray*}
so, $X^T\mathbf{c}=\mathbf{0}$.

\subsection{Conditions for consistency of the QR estimator}\label{consistencyConditions}
Let the $q$th conditional quantile function of $Y|\mathbf{X}$ be $Q_Y(q|\mathbf{X})$.
Per \citep[Section 4.1.2]{koen:2005}, for the quantile regression estimator $\hat{\bm{\beta}}_q$
to be consistent, the following conditions have to hold:
\begin{enumerate}
\item There exists $d > 0$ such that
$$\underset{n\rightarrow\infty}{\lim\inf} \underset{\|u\|=1}{\inf} n^{-1}\sum I(|\mathbf{x}_i'u|<d)=0\,.$$
\item There exists $D > 0$ such that
$$\underset{n\rightarrow\infty}{\lim\sup} \underset{\|u\|=1}{\sup} n^{-1}\sum (\mathbf{x}_i'u)^2\le D\,.$$
\end{enumerate}

\clearpage

\pagebreak
\subsection{Simulation Scenarios}
\begin{table}[h!]
\tiny{
\resizebox{\textwidth}{!}{
\begin{tabular}{|l|l|l|l|}
\hline
No. & Description & Model for the response, $y$ & Error distribution\\
\hline
1    & Intercept only & $3$ & $N(0,0.25^2)$\\
2-11 & Simple linear model & $5-x$ & $N(0,\sigma^2)$, \\
     &                     &       & $\sigma\in\{0.1,0.2,\ldots,1\}$\\
12   & Two predictors & $1-3x_1+2x_2$ & $N(0,0.1^2)$\\
13   & Five predictors & $1-3x_1+2x_2+2x_3$ & $N(0,0.1^2)$\\
     &                 &  \hspace{2cm}$-x_4-2x_5$       & \\
14   & s.d. increases linearly  & $3+2x$ & $N(0,(0.1+0.2x)^2)$\\
15   & s.d. increases linearly  & $5+x$ & $N(0,(0.1+0.5x)^2)$\\
16   & s.d. increases linearly & $3+0.5x$ & $N(0,(0.5+0.7x)^2)$\\
17   & Polynomially increasing s.d. & $1-2x$ & $N(0,(0.1+0.2x^3)^2)$\\
18   & Linearly decreasing s.d. & $7+3x$ & $N(0,(1-0.5x)^2)$\\
\hline
19   & Intercept only & $5$ & $LN(0,0.75)$\\
20   & Simple linear model & $3-x$ & $LN(0,0.75)$\\
21   & Five predictors & $1-3x_1+2x_2+2x_3$ & $LN(0,0.75)$\\
     &                 & \hspace{2cm}$-x_4-2x_5$ & \\
22   & Linearly increasing (log) s.d. & $2-2x$ & $LN(0,0.5+0.5x)$\\
\hline
23   & Quadratic, increasing variance & $6x^2+x+120$ & $N(0, (0.2+x)^2)$\\
24   & Interaction, increasing variance  & $4x_1x_2$ & $N(0, (0.1+0.2x_1)^2)$\\
\hline
25   & Mixed model  & $2 + x + z u$ & $u \sim N(0,0.5^2)$, \\
     &              &               & $e\sim N(0, 0.1^2)$\\
\hline
\end{tabular}}
\caption{Simulation scenarios. In simulations 1-11, 14-17, and 24 the predictors were drawn uniformly 
from  $(0,1)$. In simulation 12 $x_1\sim U(0,1)$ and $x_2\sim U(-3,3)$.
In simulations 13, 18-22 the predictors were drawn uniformly 
from  $(-1,1)$ and in simulation 23, from $(-5,5)$.
In simulation 25, $x_{it}\sim N(t/4, 0.1^2)$, for $t=1,2,3,4$.}\label{sims}
}
\end{table}


\begin{table}[ht!]
\tiny{
\resizebox{\textwidth}{!}{\begin{tabular}{|l|l|l|}
\hline
& Description & Model\\
\hline
1 & Intercept only, variance increases  & $Y\sim N(3,(0.1+X_1)^2)$\\
  &   with a predictor, $X_1$ & \\
2 & Simple linear regression, constant variance & $Y\sim N(5-X_1,0.3^2)$\\
3 & Simple linear regression, increasing variance & $Y\sim N(5-X_1,(0.1+X_1)^2)$\\
4 & Multiple predictors, constant variance & $Y\sim N(1-3X_1+2X_2+2X_3$\\
  &   & \hspace{1cm}$-X_4-2X_5,0.1^2)$\\
5 & Multiple predictors, increasing variance & $Y\sim N(1-3X_1+2X_2+2X_3$\\
  &   & \hspace{1cm}$-X_4-2X_5,(0.1+X_1)^2)$\\
6 & Multiple predictors,  variance  & $Y\sim N(1-3X_1+2X_2+2X_3$\\
  &   which depends on two predictors & \hspace{1cm}$-X_4-2X_5,(0.1+X_1+1.3X_2)^2)$\\
7 & Non-normal errors & $Y\sim 1-3X_1+2X_2+2X_3-X_4$\\
  &   & \hspace{1cm}$-2X_5 +LN(0,0.75^2)$\\
8 & Non-normal errors which depend on $X_1$ & $Y\sim 2-2X_1+LN(0,(0.25 + 0.5X_1)^2)$\\
9 & Normal errors, correlated variables & $Y\sim N(\sum_{i=1}^{20}X_i,0.1^2)$\\
   & &  \hspace{1cm}$cor(x_i,x_{i+1})=0.95$ $i=1,\ldots,19$\\
\hline
\end{tabular}
}
\caption{Simulation scenarios, large $P$. In the simulations with lognormal errors, the parameters are expressed on the logarithmic scale.}\label{simsLargeP}
}
\end{table}

\begin{figure}%
\centering
\subfigure[$y\sim N(5+x,(0.1+0.5x)^2)$.]{%
\label{sdbeta1sim15}%
\includegraphics[trim={0 0.5cm 2 1cm},clip, width=.45\linewidth]{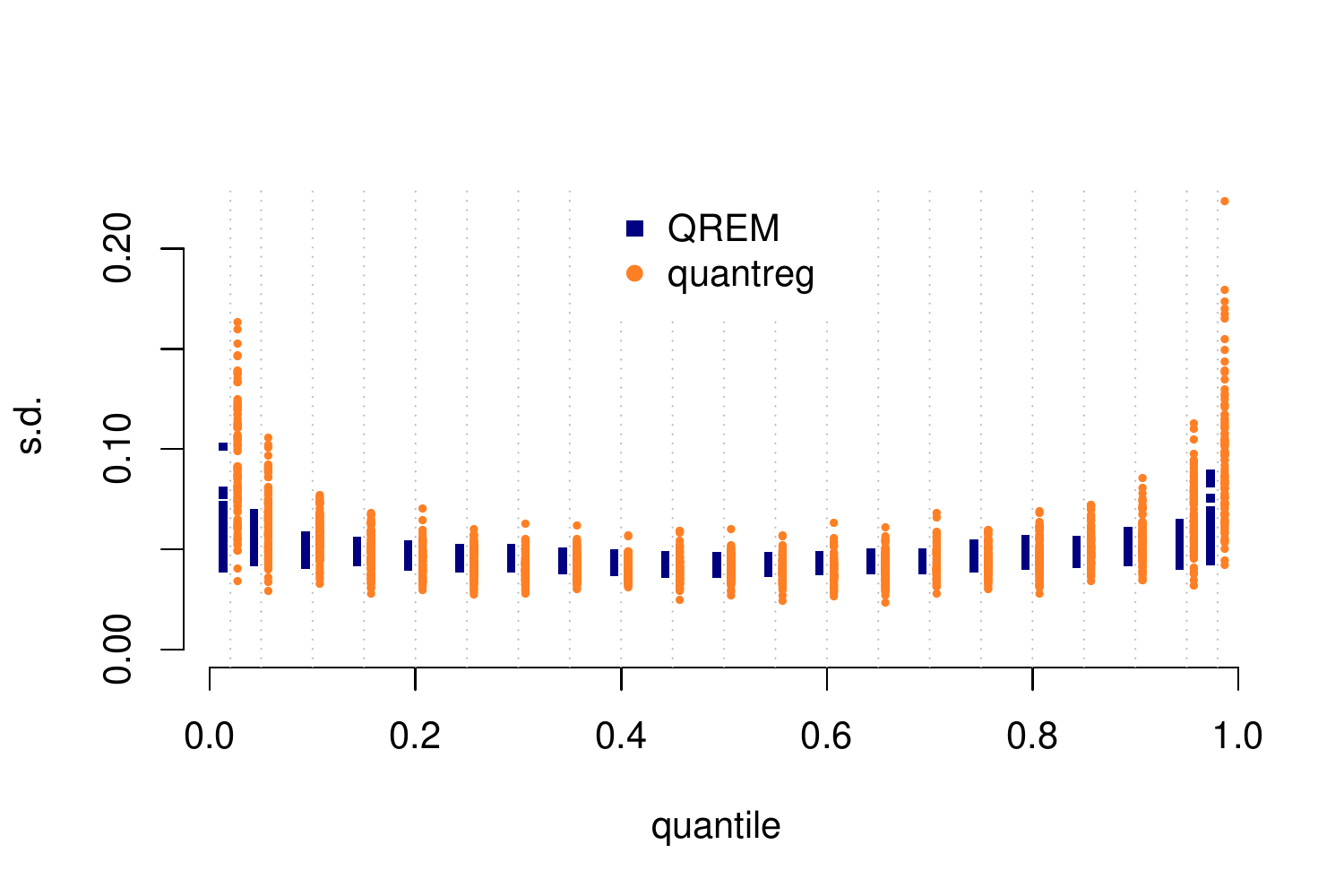}}%
\qquad
\subfigure[$y=1-3x_1+2x_2+2x_3-x_4-2x_5+\epsilon_i$ and $\epsilon_i\sim LN(0,0.75)$]{%
\label{sdbeta1sim21}%
\includegraphics[trim={0 0.5cm 2 1cm},clip, width=.45\linewidth]{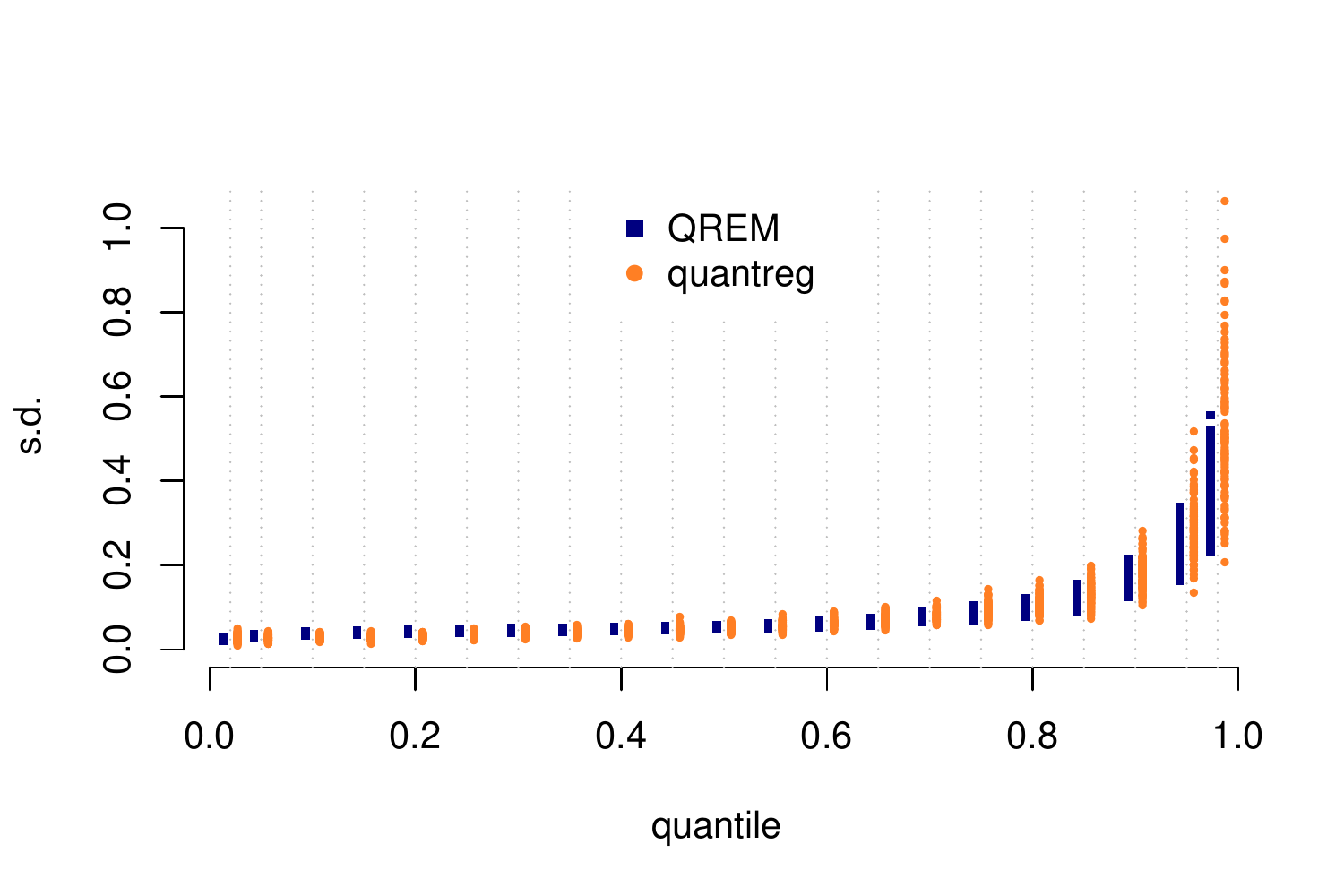}}%
\caption{QR simulations \#15 (left) and \#21 (right): $\hat\sigma_{\hat\beta_1}$ for $q\in\{0.02,0.05,0.1,\ldots,0.95,0.98\}$}
\end{figure}

\end{document}